\documentclass{aastex}
\usepackage{emulateapj5}
\slugcomment{accepted by ApJ}

\shortauthors{A. K. Inoue}
\shorttitle{Dust in H {\sc ii} Regions}

\begin{document}

\title{Spatial Distribution of Dust Grains within H {\sc ii} Regions}

\author{Akio K. INOUE}
\affil{Department of Astronomy, Faculty of Science, Kyoto University,
Sakyo-ku, Kyoto 606-8502, JAPAN}
\email{inoue@kusastro.kyoto-u.ac.jp}

\begin{abstract}

We discuss the dust distribution within photoionized regions.
Assuming a geometry with a central dust cavity, which is strongly
 suggested by the literature, we can estimate the cavity's radius from
 the ratio of the infrared and radio fluxes by using a simple transfer
 model of Lyman continuum photons.
We apply the method to a sample of the Galactic H {\sc ii} regions.
The estimated typical radius of the dust cavity of the Galactic compact
 H {\sc ii} regions is about 30\% of the Str{\"o}mgren radius.
Taking account of uncertainties both of the observational data and the
 model, we can reject a dust distribution lacking a central cavity.
Therefore, the dust cavity model is supported independently of the
 previous works.
We discuss the formation mechanism of such a dust cavity and its
 detectability by present and future infrared facilities.

\end{abstract}

\section{Introduction}

Dust grains exist everywhere.
The inside of H {\sc ii} regions is not an exception (e.g.,
\citealt{ish68, har71, mel79}).
The radiation from celestial objects is always absorbed and scattered
by the grains.
Without correction for the dust extinction, we inevitably underestimate
the intrinsic intensity of the radiation, and our understanding of
the physics of these objects may be misled.
In order to evaluate the amount of the dust extinction in detail, it
is important to know the distribution of dust as well as its
optical properties.

Theoretically, some attempts to reveal the dust distribution in H {\sc
ii} regions have been made to date \citep{mat67,mat69,gai79a,gai79b}.
According to \cite{gai79b}, the radiation pressure on the dust grains by
the central source causes a central hole of dust and gas.
The dust distribution has also been observationally investigated.
{}From the observed flux ratio of H $\beta$ to continuum, \cite{ode65}
have shown that the gas-dust ratio decreases with radius in the Orion
nebula.
{}From radial photometric profiles at $V$-band of some H {\sc ii}
regions, \cite{nak83} also suggest that a dust depletion zone is in the
central region (See also \citealt{yos86,yos87}).
Moreover, the central dust cavity is supported by
fitting to the observed infrared (IR) spectral energy distribution 
(SED) of H {\sc ii} regions \citep{chi86,chi87,chu90,fai98,gho00}.
Recently, such dust geometry is preferred not only by IR SED 
fitting but also by the radial photometric profiles at submillimeter
range \citep{hat00}.
In addition, we have some evidence of central gaseous cavities from
radio observations \citep{ter65,woo89}.
Since dust and gas are well coupled each other \citep{gai79b}, the
presence of a gaseous cavity indicates a dust cavity.

Although the dust cavity is likely to exist, it has not been observed
directly in the far-infrared (FIR) band, which is dominated by the
thermal emission from dust.
This is because the FIR observations are not advanced relative to
other wavelengths, and moreover, the spatial resolving power of FIR
imaging is generally weak.
Recently, more detailed observational data have been obtained by {\it
ISO} (Infrared Space Observatory) and SCUBA (Submillimeter Common-User
Bolometer Array), and many successive facilities are
planned in the near future (SIRTF, SOFIA, ASTRO-F, ALMA, etc.).
By using these new powerful facilities, we may observe the dust
distribution directly.
We need to estimate the detectability of the cavity by these facilities.

On the other hand, we have dveloped a method for determining the
fraction of Lyman continuum (LC) photons contributing to hydrogen
ionization (\citealt{ihk01, ino01}, hereafter Paper I and Paper II,
respectively).
In paper I and II, we show that only half of LC photons from the central 
source in an H {\sc ii} region ionizes neutral hydrogens, and the rest
are absorbed by dust grains within the ionized region.
If a dust cavity exists in the ionized region, however, the efficiency
of the dust absorption for the LC photons becomes small.
In this paper, we examine whether the estimated photon fraction is
reproduced  by a model nebula with a central dust cavity.

In the next section, we describe our method for examining the dust
distribution, especially the central dust cavity's radius, by using the
fraction of LC photons contributing to hydrogen ionization.
In section 3, we apply it to the sample H {\sc ii} regions in the
Galaxy.
Then, we discuss the formation mechanism of the dust cavity and its
detectability by some IR facilities in section 4.
Finally, we summarize our conclusions in the last section.

\section{Method}

\subsection{Photon Transfer}

We assume an ideal spherical nebula composed of only hydrogen and dust
because the effect of helium is not so important for the determination
of the dust distribution.
This will be discussed in more detail in the last paragraph of section
2.3.2.
Moreover, we assume the volume number densities of hydrogen and dust
to be constant, $n_{\rm H}$ and $n_{\rm d}$, respectively.
If densities of neutral hydrogen atoms, protons, and electrons, are
represented by $n_{\rm H^0}$, $n_{\rm p}$, and $n_{\rm e}$,
respectively, we have $n_{\rm H^0}=(1-x)n_{\rm H}$ and $n_{\rm p}=n_{\rm
e}=xn_{\rm H}$ in terms of the ionization degree, $x$, which varies with
the radius, $r$, from the central exciting source.
In addition, we assume the frequencies of all LC photons to be that of
the Lyman limit, that is, all quantities are treated monochromatically
in our transfer model.
This is because we obtain only the number of LC photons from
observations of recombination lines or thermal radio emissions from H
{\sc ii} regions.

We introduce a dimensionless radius, $y=r/R_{\rm S}$, where $R_{\rm S}$
is the classical Str{\"o}mgren radius which is defined as 
\begin{equation}
 R_{\rm S} = \left(\frac{3N_{\rm LC}}{4\pi n_{\rm H}^2 \alpha_B}
             \right)^{1/3}\,.
 \label{eq2-0}
\end{equation}
Here $N_{\rm LC}$ is the intrinsic production rate of LC photons of the
central source, and $\alpha_B$ is the Case B recombination
coefficient of hydrogen.

We consider the change of the number of LC photons passing
through the shell whose inner and outer radii are $y$ and $y+dy$,
respectively.
If the number of LC photons entering the shell per unit time denotes
$N_{\rm LC}(y)$, its change in the shell is 
\begin{eqnarray}
 dN_{\rm LC}(y)= 
 - N_{\rm LC}(y)\{n_{\rm H^0}s_{\rm H}+(1-\omega)n_{\rm d}s_{\rm d}\} 
 R_{\rm S} dy \nonumber \\
 + 4\pi R_{\rm S}^3 y^2 n_{\rm p} n_{\rm e} \alpha_1 dy\,,
 \label{eq2-1}
\end{eqnarray}
where $s_{\rm H}$ and $s_{\rm d}$ are cross sections for LC 
photons of neutral hydrogen atoms and dust grains, respectively,
$\omega$ is the dust albedo for these photons, and $\alpha_1$ is the
recombination coefficient to the ground state of hydrogen.
We neglect the asymmetric effect of the dust scattering for simplicity.

Since the shell is in ionization equilibrium, 
\begin{equation}
 N_{\rm LC}(y) n_{\rm H^0} s_{\rm H} R_{\rm S} dy 
 = 4 \pi R_{\rm S}^3 y^2 n_{\rm p} n_{\rm e} \alpha_A dy\,,
 \label{eq2-2}
\end{equation}
where $\alpha_A$ is the coefficient of hydrogen recombination to all
levels.
We note that $\alpha_A-\alpha_B=\alpha_1$.
By equation (\ref{eq2-2}) and replacing $n_{\rm H^0}$ with
$(1-x)n_{\rm H}$, equation (\ref{eq2-1}) is reduced to
\begin{equation}
 dN_{\rm LC}(y) = - N_{\rm LC}(y) \{(1-x)\frac{\alpha_B}{\alpha_A} + X \}
  n_{\rm H} s_{\rm H} R_{\rm S} dy\,,
 \label{eq2-4}
\end{equation}
where 
\begin{equation}
 X = (1-\omega) \frac{n_{\rm d}s_{\rm d}}{n_{\rm H}s_{\rm H}}\,.
 \label{eq2-5}
\end{equation}
This is a correction term for existence of dust grains in the nebula.
If we define the total (gas + dust) optical depth for LC photons as 
\begin{equation}
 d\tau = \{(1-x)\frac{\alpha_B}{\alpha_A} + X\} 
 n_{\rm H} s_{\rm H} R_{\rm S} dy\,, 
 \label{eq2-5.1}
\end{equation}
we obtain $N_{\rm LC}(y)=N_{\rm LC}\,e^{-\tau}$.
Then equation (\ref{eq2-2}) is reduced to 
\begin{equation}
 \frac{1-x}{x^2} = \frac{4\pi R_{\rm S}^2 y^2 n_{\rm H} \alpha_A}
                        {s_{\rm H} N_{\rm LC} \, e^{-\tau}}\,.
 \label{eq2-3}
\end{equation}

Now we define the ionization parameter as 
\begin{equation}
 U = \frac{N_{\rm LC}}{4\pi R_{\rm S}^2 n_{\rm H} c}
   = \frac{1}{c}
     \left(\frac{{\alpha_B}^2 n_{\rm H} N_{\rm LC}}{36\pi}\right)^{1/3}\,,
 \label{eq2-5.2}
\end{equation}
where $c$ is the light speed.
Then equations (\ref{eq2-5.1}) and (\ref{eq2-3}) are reduced to
\begin{equation}
 \frac{d\tau}{dy} = \frac{3cs_{\rm H}}{\alpha_B}
 \{(1-x)\frac{\alpha_B}{\alpha_A} + X\} U\,,
 \label{eq2-5.3}
\end{equation}
and
\begin{equation}
 \frac{1-x}{x^2} = \frac{\alpha_A y^2 e^\tau}{s_{\rm H} c U}\,,
 \label{eq2-5.4}
\end{equation}
respectively.

Finally, we determine two recombination coefficients of hydrogen,
$\alpha_A$ and $\alpha_B$.
The coefficients are a function of the electron temperature, $T_{\rm
e}$ \citep{spi78}.
Fitting values in Table 5.2 of \cite{spi78}, we can approximate them to
the following equation;
\begin{equation}
 \frac{\alpha_i}{\rm cm^3\,s^{-1}} = 2.06\times10^{-11}
  \frac{a_i - \log{T_{\rm e}/4000\,{\rm K}}}{\sqrt{T_{\rm e}/{\rm K}}}\,,
 \label{eq2-5.5}
\end{equation}
for $2000\,{\rm K} \la T_{\rm e} \la 16000\,{\rm K}$, where $i=A$ or
$B$, and $a_A=2.40$ and $a_B=1.64$.
The uncertainties of values determined by these equation are less than
1\% for the above range of $T_{\rm e}$.

\subsection{Dust Properties}

Let us estimate the dust effect that appears in the term, $X$.
We adopt the dust extinction law in the Galactic interstellar medium
(ISM) developed by \cite{wei01}, who have derived a size-distribution
function of dust grains composed of carbonaceous and silicate
populations by \cite{dl84} and \cite{li01}.
Their carbonaceous grain includes PAH (polycyclic aromatic hydrocarbon)
molecules.
\cite{wei01} have developed the extinction laws for various values of
$R_V$, which is the ratio of the visual extinction to color excess.

We approximate the optical depths for LC photons by hydrogen and dust to
those at 912 \AA.
According to \cite{wei01}, the dust cross sections per unit hydrogen
nucleus at 912 \AA\ i.e. $n_{\rm d}s_{\rm d}^{\rm 912\AA}/n_{\rm H}$ are
$2.515 \times 10^{-21}$ cm$^2$, $1.676 \times 10^{-21}$ cm$^2$, and
$1.087 \times 10^{-21}$ cm$^2$, for $R_V=3.1$, 4.0, and 5.5,
respectively.\footnote{We adopt the extinction laws with the 
parameter sets recommended by \cite{wei01}. For $R_V=3.1$, we adopt the
law with the parameter set of Case A and $b_{\rm C}=6.0\times10^{-5}$,
which represents C abundance of very small grain population. Also, we
adopt the laws with the parameter sets of Case B and $b_{\rm
C}=4.0\times10^{-5}$, or Case B and $b_{\rm C}=3.0\times10^{-5}$ for
$R_V=4.0$, or 5.5, respectively. Values of other parameters are
tabulated in Table 1 in \cite{wei01}.}
We note that these values are calibrated by $A_I/N_{\rm
H}=2.6\times10^{-22}$ cm$^2$, where $N_{\rm H}$ is the hydrogen column
density.
Also, the dust albedos at 912 \AA\ are 0.2451, 0.2694, and 0.2989 for
$R_V=3.1$, 4.0, and 5.5, respectively.
If we adopt $s_{\rm H}^{\rm 912\AA} = 6.3\times10^{-18}$ cm$^2$,
equation (\ref{eq2-5}) is reduced to 
\begin{equation}
 X = \cases{3.0\times10^{-4} & for $R_V=3.1$\cr
            1.9\times10^{-4} & for $R_V=4.0$\cr
	    1.2\times10^{-4} & for $R_V=5.5$}\,.
 \label{eq2-6}
\end{equation}

Here, we introduce a parameter for the dust distribution, the
dimensionless dust cavity's radius, $y_{\rm d}$, because the dust
distribution may have a radial dependence as strongly suggested in the
literature (\S 1).
For simplicity, we consider an extreme case that no dust is inside
the cavity's radius, $y_{\rm d}$, and the gas-dust ratio is constant
outside of it.
That is, the dust term, $X$, is zero within $y_{\rm d}$, and constant
value shown above outside of it.

Now, we can solve the photon transfer problem.
If the ionization parameter, $U$, is given, and $\tau=0$ and $x=1$
at $y=0$ are adopted as the boundary condition, we obtain the ionization
degree, $x$, at any radius, $y$, from equations (\ref{eq2-5.3}) and
(\ref{eq2-5.4}) for a set of $R_V$ and $T_{\rm e}$.
The dust inner radius, $y_{\rm d}$, is only one free parameter in our
photon transfer model.

\subsection{Ionizing Photon Fraction}

As investigated in Paper I and II, not all of the intrinsic LC photons
emitted by the central source of H {\sc ii} regions is consumed in
hydrogen ionization.
In other words, the number of LC photons contributing to the ionization,
i.e. ionizing photons, is always less than the intrinsic one.
Then, we define the ionizing photon fraction, $f$, as  
\begin{equation}
 f = \frac{N'_{\rm LC}}{N_{\rm LC}}\,,
 \label{eq2-6.1}
\end{equation}
where $N'_{\rm LC}$ and $N_{\rm LC}$ are the production rates of the
ionizing and intrinsic LC photons, respectively.
For the ionizing photons, the production rate, $N'_{\rm LC}$, is
determined from the observational data, whereas the intrinsic production
rate, $N_{\rm LC}$, cannot be obtained observationally.
Let us determine the fraction, $f$, via two ways below.

\subsubsection{From Transfer Model}

First, we determine the fraction from our transfer model.
According to Paper I, the fraction, $f$, is reduced to $y_{\rm i}^3$ for
a perfectly ionized nebula (i.e., $x=1$ for $y \le y_{\rm i}$ and $x=0$
for $y > y_{\rm i}$), where $y_{\rm i}$ is the dimensionless radius of
the ionized region.

Total number of proton in a nebula is $\int x n_{\rm H} 4\pi R_{\rm S}^3 
y^2 dy$ in our transfer model.
On the other hand, the proton number is equal to $(4\pi/3) R_{\rm S}^3
y_{\rm i}^3 n_{\rm H}$ in the perfectly ionized nebula.
Since our solutions of the photon transfer model indicate that the
nebula is regarded as the perfectly ionized nebula, we can equate above
two quantities each other.
Thus, we obtain 
\begin{equation}
 f = 3 \int_0^{y_{\rm max}} x y^2 dy\,,
 \label{eq2-7}
\end{equation}
where $y_{\rm max}$ is the position to stop calculating.
We can choose the value of $y_{\rm max}$ arbitrarily if it is larger
than $y_{\rm i}$.
In this paper, $y_{\rm max} = 1.2$ is adopted as a constant for simplicity.
Therefore, we can determine $f$ as a function of the dust inner
radius, $y_{\rm d}$, for three $R_V$ values, if an ionization parameter,
$U$, is given.
We note that the fraction determined from above equation
(\ref{eq2-7}) becomes equivalent to that in section 2 of Paper I if 
$y_{\rm d} = 0$ and $\omega = 0$.

In Figure 1, we show such results for various sets of $U$ and $R_V$ when 
we adopt $T_{\rm e}=6600$ K which is a mean electron temperature of the
sample H {\sc ii} regions described in section 3.
We find that $f$ decreases if $y_{\rm d}$ decreases.
This is because the dust optical depth is large as the cavity's radius
is small.
We also find that $f$ decreases if $U$ increases, which is equivalent
that $f$ decreases if $N_{\rm LC}$ or $n_{\rm H}$ increases
(eq.[\ref{eq2-5.2}]).
This indicates that the ionizing photon fraction for a dust-gas
ratio\footnote{In this paper, we adopt a dust-gas ratio corresponding to 
$A_I/N_{\rm H}=2.6\times10^{-22}$ cm$^{2}$ (see section 2.2). Since the
dust-gas ratios are different from one region to others (e.g.,
\citealt{sta00} for the SMC), we should determine the ratio for the
individual H {\sc ii} region. However, we use the above dust-gas ratio
for all sample regions, because we cannot determine it individually.}
becomes smaller as the number of LC photons from the central source
becomes larger or as the density of the surrounding medium becomes
higher.

The relation between $f$ and $y_{\rm d}$ is also a function of $R_V$, as 
shown especially in Figure 2 (a).
That is, the ionizing photon fraction for a fixed dust cavity's radius
becomes smaller, as the value of $R_V$ becomes smaller.
This is because the dust extinction law of a smaller $R_V$ shows a
steeper rise at the far-ultraviolet range \citep{wei01}.
Since a higher value of $R_V$ is suitable for a higher density region
\citep{car89}, we select $R_V=5.5$ for our sample H {\sc ii} regions
discussed in section 3.
Moreover, we see the variation of the $f$-$y_{\rm d}$ relation for the
change of the adopted electron temperature, $T_{\rm e}$.
We find in Figure 2 (b) that $f$ becomes smaller as $T_{\rm e}$ becomes
higher, because recombination coefficients becomes larger as $T_{\rm e}$
increases.

\subsubsection{From Observational data}

Next, we determine the ionizing photon fraction, $f$, from the
observational data.
According to Paper I and II, we can estimate $f$ from the 
ratio of the IR luminosity, $L_{\rm IR}$, to the ionizing photon
production rate, $N'_{\rm LC}$, 
(equation [2] in Paper II) unless a significant amount of LC photons
escapes from H {\sc ii} regions.\footnote{For details of the
derivation, see also \cite{ihk00}.}
However, in the derivation in Paper II, the stellar SED are
assumed to be the Planck function.
Thus, we overestimate the intrinsic production rate of LC photons,
and then, we underestimate $f$ in Paper II.

In this paper, we adopt a realistic stellar spectra, the Kurucz's
ATLAS 9 spectra with solar abundance and turbulence speed of 2 km
s$^{-1}$.
Then, if a Salpeter's IMF (0.1--100 $M_\sun$) is assumed, we derive the
following formula;
\begin{equation}
 f = \frac{0.375 + 0.625\epsilon}
          {0.250 + 4.42 L_{\rm IR,6}/N'_{\rm LC,49}}\,,
 \label{eq2-8}
\end{equation}
where $\epsilon$ denotes the average efficiency of dust absorption for
nonionizing UV ($\lambda>912$ \AA) photons, and $L_{\rm IR,6}$ and
$N'_{\rm LC,49}$ are the total IR luminosity and ionizing 
photon production rate normalized by $10^6$ $L_\sun$ and $10^{49}$
s$^{-1}$, respectively.
The value of $f$ obtained from the above equation is typically 1.2 times
larger than that from the equation (2) in Paper II because of the
modification of adopted stellar SED.
When we estimate $\epsilon$ from the observed color excess (see Paper II 
for detail), we determine $f$ of individual H {\sc ii} region from
its $L_{\rm IR}$ and $N'_{\rm LC}$.
It is worthwhile to nate that since the above equation is based on a
simple energy conservation, the obtained $f$ is independent of the dust
distribution, the structure of H {\sc ii} regions and so on.

Once $f$ is obtained, we can determine the intrinsic LC photon
production rate by using equation (\ref{eq2-6.1}), and can determine the 
ionization parameter, $U$.
Using the obtained $U$, we determine $y_{\rm d}$ so as to reproduce
$f$ determined from equation (\ref{eq2-8}) by using our photon transfer
model (i.e. eq.[\ref{eq2-7}]).
In the next section, we determine a typical $y_{\rm d}$ for a sample of
the Galactic H {\sc ii} regions.

We discuss the effect of the presence of helium.
Of course, helium atoms absorb LC photons as well as hydrogen and dust.
However, the fraction of stellar LC photons ionizing helium is small,
$\la$ 10\%.
Even if all helium ionizing photons are consumed by only helium, the
fraction of photons ionizing hydrogen, $f$, becomes at most 10\% smaller
than that determined by equation (\ref{eq2-8}).
Thus, the determined $y_{\rm d}$ decreases at most 10\%.
Since not all helium ionizing photons are absorbed by helium and most of
helium recombination photons can ionize hydrogen \citep{mat71}, the
effect of helium is negligible for our analysis.

\section{Results}

In order to determine the inner radius of the dust distribution, $y_{\rm 
d}$ of an H {\sc ii} region, we need various data; the electron
density and temperature, the ionizing photon production rate, the
infrared luminosity, and $R_V$.
Here, the value of $R_V$ is assumed to be constant, 5.5.
The effect of the change of $R_V$ is discussed quantitatively in the
last part of this section.

We find a suitable data set of the Galactic H {\sc ii} regions in
\cite{sim95}, which examined the physical properties of 23 H {\sc ii}
regions in the Galaxy observed by the Kuiper Airborne Observatory.
We select 13 H {\sc ii} regions from the sample of \cite{sim95}.
The adopted selection criterion is that the measured electron number
density exceeds 1000 cm$^{-3}$.
The selected regions are classified into the 'compact' or
'ultra-compact' H {\sc ii} regions from their densities \citep{hab79}.
The spherical geometry of our transfer model may be not so bad for such
sample, although even the 'ultra-compact' H {\sc ii} regions sometimes
show non-spherical shapes \citep{woo89}.
For more evolving H {\sc ii} regions than the 'compact' class, which
often show very complex shapes, e.g., 'blister', 'bubble', 'champagne',
etc., the spherical assumption is not valid.
The observed and measured quantities of the sample regions are
tabulated in Table 1.

First, we estimate the ionizing photon fraction, $f$, for each sample
region.
The ionizing photon production rate, $N'_{\rm LC}$, is estimated from
the flux density of free-free radio emission at 5 GHz by the formula in
\cite{con92}: $N'_{\rm LC}/{\rm s}^{-1}=8.88 \times 10^{46} (T_{\rm
e}/10^4{\rm K})^{-0.45} (\nu/5{\rm GHz})^{0.1} (D/{\rm kpc})^2
(S_\nu/{\rm Jy})$.
The total IR (8--1000 \micron) luminosity, $L_{\rm IR}$, is estimated
from the observed IR (18--160 \micron) flux by an upward factor of 1.26, 
assuming the dust SED to be the modified Planck function of 30 K with
the emissivity index of 1.
The conversion factor varies between 1.03--1.26 if we change the adopted 
parameter sets of the dust temperature (30--40 K) and the emissivity
index (1--2).
Thus, the conversion causes at most 20 \% uncertainty of the derived IR
luminosity.
The determined $N'_{\rm LC}$ and $L_{\rm IR}$ are given in columns (2)
and (3) of Table 2, respectively.

Since we do not have the color excess, $E_{B-V}$, of sample regions, we
assume $\epsilon = 1$, which correspond to a typical $E_{B-V}$ of the
Galactic H {\sc ii} regions observed by \cite{cap00} (See Paper II).
The determined $f$ may be overestimated if the true value of
$\epsilon$ is smaller than unity.
Indeed, if we set $\epsilon=0.8$, the value of $f$ becomes about 90 \%
of the determined here.
Now, we can determine $f$ for each sample region via equation
(\ref{eq2-8}), and give the values in column (4) of Table 2.
The mean value and standard deviation of $f$ for 13 sample regions are
0.55 and 0.22, respectively.
Thus, only half of LC photons contributes to hydrogen ionization in
the H {\sc ii} regions of the Galaxy.
This is consistent with the results in \cite{aan78}, Paper I and II.

Since the ionizing photon fraction, $f$, is indirect measurement value,
many uncertainties of adopted quantities accumulate.
Suppose a quantity, $\eta$, is represented by a functional form,
$\phi(x_1,\,x_2,\,...)$, where $x_i$ is more basic quantities.
If uncertainties of $x_i$ are represented by $\delta_i$, and all
quantities of $x_i$ are independent each other, the uncertainty of $\eta$
can be expressed by $\sum (\partial \phi / \partial x_i)^2 \delta_i$.
In this way, the uncertainty of $L_{\rm IR}/N'_{\rm LC}$ is estimated to
be about 30 \% typically.
The main causes of the uncertainty are that of the adopted radio flux
densities, about 20 \% \citep{sim95}.
Taking account of the uncertainties of the coefficients in equation
(\ref{eq2-8}) and the parameter, $\epsilon$, we estimate a typical
uncertainty of $f$ to be about 33 \%, which corresponds to $\pm0.2$ for
the mean $f$.
Thus, a somewhat large value of $f$ for G298.22-0.34, in stead of $f \leq
1$ by the definition, may be caused by this uncertainty.

Next, we determine the dust cavity's radius, $y_{\rm d}$, of sample
regions.
By using the estimated $f$, we can convert the ionizing photon
production rate, $N'_{\rm LC}$, into the intrinsic LC one, $N_{\rm
LC}$, that is, $N_{\rm LC} = (1/f) N'_{\rm LC}$.
Then, we calculate the ionization parameter, $U$, of the individual
sample region by equation (\ref{eq2-5.2}) shown in column (5) in
Table 2.
Here, we determine $y_{\rm d}$ so that $f$ estimated from equation
(\ref{eq2-8}) is equal to $f$ calculated from our transfer model (i.e.,
eq. [\ref{eq2-7}]).\footnote{Determining $y_{\rm d}$, we adopt the
recombination coefficients calculated from the individual electron
temperature by equation (\ref{eq2-5.5}).}
The obtained values of $y_{\rm d}$ are given in the last column in Table
2.
The mean value of the determined $y_{\rm d}$ and its standard deviation 
are 0.39 and 0.23, respectively.
The uncertainty of the individual value of $y_{\rm d}$ caused by that of
the estimated $f$ is about $\pm0.2$.

In Figure 3, we show histograms of the derived quantities.
{}From the panels (c) and (d), we can divide the sample into two
groups: one is the 'ultra-compact' H {\sc ii} regions, whose Str{\"o}mgren 
radius, (or the ionized radius, $r_{\rm i}$) is smaller than 0.5 pc (0.4
pc).
The other is the 'compact' H {\sc ii} regions with more larger radii.
Here, we should note that G298.22-0.34 is removed from the following
discussions because of its anomalous value of $f$, and S156 is also
removed because of its very small luminosity relative to others.
The two classifications are indicated in Figure 3 and Table 1.
In Table 3, we summarize the mean properties of the sample of the two
groups.
Although our classification is based on the radius of sample regions,
the classification is caused by the difference of the electron density,
and are consistent with that of \cite{hab79}.

For the 'compact' sample, we estimate the mean value of $y_{\rm d}$ to
be $0.30\pm0.12$, where the error is the sum of two uncertainties of the 
mean value; one results from the variation among sample regions, and the
other from the uncertainty of individual value.
The estimated error is coincident with the sample standard deviation.
For the 'ultra-compact' sample, the mean value of $y_{\rm d}$ is
estimated to be $0.48\pm0.26$.
The error is the same mean as that of the 'compact' sample.

The non-zero mean value of $y_{\rm d}$ obtained above may results from a
somewhat large uncertainty of individual object.
If we assume that the sample regions compose a normal population, 
we can test whether the mean $y_{\rm d}$ of the population is regarded
to be zero.
If the variance of $y_{\rm d}$ for the population is estimated to be
square of the uncertainty of individual value, $(0.2)^2$, we can reject 
the hypothesis of the mean $y_{\rm d}=0$ with the significance level of
0.0011\% for the 'compact' sample, and 0.0016\% for the 'ultra-compact'
sample.
Therefore, even if we take account of somewhat large uncertainties of
the observations and the model, the ionizing photon fraction cannot be
reproduced by a nebula filled with dust uniformly (i.e. $y_{\rm d}=0$).
The dust distribution with the central cavity is supported by our
analysis independently of the IR SED fitting or the photometric profile
analysis.

The cavity's radius in units of the Str{\"o}mgren radius (i.e. $y_{\rm
d}$) of the 'ultra-compact' sample is larger than that of the 'compact'
sample.
The real scale of the cavity's radii of both samples are about 0.3 pc
and 0.2 pc for 'compact' and 'ultra-compact', respectively.
We may see an evolutionary sequence of the dust cavity from
'ultra-compact' to 'compact'.
However, these values are agree with each other within their error
bars, and the 'ultra-compact' sample consists of only three regions.
Thus, it is uncertain whether the difference between the cavity's radii
of two subsamples is real or not.

Finally, we demonstrate the effect of the change of $R_V$ on the
obtained dust cavity's radius, $y_{\rm d}$.
As shown in Figure 2 (a), the value of $y_{\rm d}$ increases for a fixed 
ionizing photon fraction, $f$, if $R_V$ decreases.
This is shown quantitatively for three sample set in Table 4.
For the 'compact' sample, the value of $y_{\rm d}$ of $R_V=3.1$ is a
factor of 1.7 times larger than that of $R_V=5.5$.

\section{Discussions}

As described above, assuming spherical symmetry and uniform density
distribution, we obtain a typical radius of the central dust cavity in
the 'compact' H {\sc ii} regions, 0.3 times Str{\"o}mgren radius.
This corresponds about 40 \% of the ionized radius and about 0.3 pc.
We discuss the formation mechanism and the detectability of the cavity.
However, we should keep in mind that since real H {\sc ii} regions, even
compact or ultra-compact regions, show a complex structure, the radius
of a real cavity may be different from 0.3 pc obtained by our model.

\subsection{Formation Mechanism}

Three formation mechanisms of the central dust cavity are naturally
expected: (1) radiation pressure, (2) stellar wind by the central
source, and (3) dust sublimation.
Here we discuss whether the above mechanisms can produce the cavity and
which is more effective.

The effect of radiation pressure of the central source on the
distribution of gas and dust in H {\sc ii} regions is discussed in
detail by \cite{gai79b}.
According to them, the radiation force acting on the dust grain produces 
the central cavity of dust and gas (See their Fig.7).
The radius of the cavity is about 20 \% of the ionized radius.
Although we cannot directly compare their cavity's radius with that in
this paper because of the difference between the adopted parameter sets, 
the radiation pressure can produce a large cavity.

Also, strong stellar wind of the massive stars must contribute to
produce the central cavity.
Indeed, the central low density cavity of gas is produced by the stellar
wind (e.g., \citealt{com97}).
Since dust and gas are well coupled with each other \citep{gai79b}, the
gas cavity indicates the dust cavity.
Unfortunately, we cannot estimate quantitatively the contribution of the 
stellar wind to form the dust cavity here.
It is beyond the scope of this paper.
We will resolve which of radiation pressure and stellar wind is dominant 
mechanism to form the cavity in our future work.

How about the effect of dust sublimation?
Indeed, dust grains may sublime in H {\sc ii} regions because of the
strong radiation from the central source.
However, the radius of the dust cavity caused by the dust sublimation is 
about $10^{-4}$ pc \citep{moo99}.
This is too small to produce the large cavity expected here.
Thus, the dust sublimation is not dominant mechanism to form the dust
cavity.

In addition, we discuss the effect of the clumpy distribution of dust.
Under the clumpy distribution, which is more realistic than the smooth
one adopted here, we may reproduce the ionizing photon fraction
determined from observational data even if the dust clumps exist in the
central area of H {\sc ii} regions.
However we consider that there is also the central cavity under the
clumpy distribution.
This is because the radiation pressure and stellar wind will blows out
the dust clumps from the central area.
We note that the expected radius of the cavity in clumpy medium may
become smaller than the determined one in the previous section.
Such effect of the clumpiness should be clarified in our future work.

\subsection{Detectability of Dust Cavity}

As stated by \cite{wri73}, the best observational test of the dust
cavity model is FIR (or submillimeter) observations with high angular
resolution.
Assuming a spherical and uniform structure, we expect that the dust
cavity's radius is about 0.3 pc for our sample of the Galactic compact H
{\sc ii} regions.
It corresponds to about $60''/(D/{\rm kpc})$, where $D$ is the distance
to the region from us.
Here, we examine the detectability of such dust cavity of 'compact' or
'ultra-compact' H {\sc ii} regions by the present and future IR
facilities.

The future IR observational satellite of Japan, ASTRO-F ({\it IRIS};
Infrared Imaging Surveyor) is planned to launch in 2004 by the Institute
of Space and Astronautical Science of Japan.
It will offer modest angular resolutions of $30''$--$50''$ at 50--200
\micron\ \citep{mur98}.
We may detect the dust cavity of H {\sc ii} regions located within only
about 1 kpc from us.
However, ASTRO-F will survey the whole of the sky, fortunately.
Thus, we may detect many such closest cavities.

SIRTF, which is the major upcoming IR satellite planned to launch in
July 2002 by NASA, offers much higher angular resolutions of
$2.5''$--$16''$ at 24--160 \micron.\footnote{See
http://sirtf.caltech.edu/SSC/}
If we observe the emission from dust being thermally equilibrium with
ambient gas, the super resolution mode at 70 \micron, which provides us
with $4.9''$ angular resolution, is the most suitable.
By this observing mode, we can detect the dust cavity of H {\sc ii}
regions within about 12 kpc from us, i.e. almost all H {\sc ii}
regions in our Galaxy.
Therefore, SIRTF will resolve the dust cavity, and clarify the dust
distribution in H {\sc ii} regions in detail.

We may also detect the dust cavity at submillimeter. 
For example, SCUBA provides us with much higher angular resolution, $8''$
at 450 \micron, or $14.5''$ at 850 \micron.
\cite{hat00} observed the Galactic 'ultra-compact' H {\sc ii} regions by
SCUBA, and present high resolution images of them.
\cite{hun00} also present high resolution images of the 'ultra-compact'
H {\sc ii} regions obtained by Caltech Submillimeter Observatory.
However, we cannot find the evidence of the cavity from their images.
This may be because the sample regions are 'ultra-compact', that is,
their densities are very high, so that the radius of the cavity is very
small.
Indeed, the estimated gas densities of their sample exceed about $10^4$
cm$^{-3}$ \citep{hun00}.
For such high density region, the Str{\"o}mgren radius is very small,
less than 0.1 pc.
It is corresponds to $\la$ $5''$ for the typical distance of their sample
(about 5 kpc).
Since the cavity's radius is less than the Str{\"o}mgren radius, the
cavity cannot be resolved by their observations.
If we select some nearby spherical 'compact' H {\sc ii} regions, whose
gas densities are about $10^3$ cm$^{-3}$, as the targets, 
we may detect the dust cavity by SCUBA.

\section{Conclusions}

We examine a possible distribution of dust grains in H {\sc ii} regions.
Assuming the geometry with the central dust cavity, which is suggested
theoretically and observationally in the literature, we can determine
the cavity's radius by using a simple transfer model of Lyman continuum
photons in a dusty spherical medium.
We adopt the extinction law recently developed by \cite{wei01}, which is
based on two components of grains, carbonaceous and silicate grains, and
is a function of the ratio of visual extinction to color excess, $R_V$.
We assume $R_V=5.5$, which is suitable for high density regions.
In the transfer model, only one free parameter is included; the radius
of the dust cavity.

The fraction of Lyman continuum photons contributing to hydrogen
ionization (i.e. ionizing photons) is adopted as a new constraint to
determine the cavity's radius.
The ionizing photon fractions of 13 'compact' or 'ultra-compact' H {\sc
ii} regions in the Galaxy are determined from the ratio of infrared to
radio fluxes via the method developed by us recently (Paper I and II).
Its mean is 0.55.
It is consistent with our previous results; a half of LC photons is
absorbed by dust before they ionize the neutral hydrogen.

We examine the effect of various quantities on the relation between
the ionizing photon fraction and the dust cavity's radius.
Setting a fixed radius of the cavity, we find that the ionizing photon
fraction decreases as the ionization parameter increases.
Also we find that the fraction decreases as the value of $R_V$ decreases 
or as electron temperature increases.

We determine the dust cavity's radius of individual sample region
so as to reproduce the fraction of ionizing photons estimated from
observational data by the transfer model.
The mean determined radius of the dust cavity is 0.39 times 
Str{\"o}mgren radius for all of the sample.
We divide our sample into two subsamples: one is the 'ultra-compact'
sample, and the other is the 'compact' sample.
Our classification is based on the Str{\"o}mgren (or ionized) radius of
the sample regions, and is consistent with that of \cite{hab79} based on
the gas densities.
A typical radius of the dust cavity for the 'compact' sample is
$0.30\pm0.12$ in units of the Str{\"o}mgren radius, where the error
includes both uncertainties of the individual estimate and the variation
among sample regions.
This corresponds to about 40\% of the ionized radius and about 0.3 pc.
Also, a typical cavity's radius for the 'ultra-compact' sample is
$0.48\pm0.26$ in units of the the Str{\"o}mgren radius, which
corresponds to about 60\% of the ionized radius and about 0.2 pc.

Since the uncertainties of these values are somewhat large and the
'ultra-compact' sample consists of only three regions, it is uncertain
whether the difference of the cavity's radius between two subsamples is
real and indicates an evolutionary sequence of H {\sc ii} regions.
In any case, we can reject the dust distribution without the central
cavity with the significance level about 0.001\% for both samples.
Thus, we conclude that the model nebula filled with dust
uniformly is invalid in order to explain the ionizing photon fraction.
The dust cavity model is supported independently of the previous works.

We discuss the formation mechanism of the dust cavity.
Although the radiation pressure and stellar wind by the central source
can produce the central cavity, it is uncertain which is dominant
mechanism.
On the other hand, the dust sublimation process is not a
major mechanism to form the cavity.
This is because the expected cavity's radius by the sublimation process
is too small to explain that obtained by us.

We discuss the detectability of the dust cavity by present and
future infrared--submillimeter facilities.
SIRTF is the most powerful facility to detect the cavity.
Indeed, we expect that SIRTF can detect the cavities in almost all H
{\sc ii} regions of our Galaxy.
SCUBA can also detect the cavities if we select nearby (typically $\la$
5 kpc) spherical 'compact' H {\sc ii} regions, whose gas densities are
about $10^3$ cm$^{-3}$, as the targets.
Moreover, Japanese IR satellite, ASTRO-F can detect the cavities in H
{\sc ii} regions located within only 1 kpc from us. 
Since ASTRO-F surveys the whole of the sky, we may detect a lot of
closest H {\sc ii} regions with the central cavity.

\acknowledgments

The author would like to thank anonymous referee for many useful comments 
to improve this paper extensively, and H. Kamaya and H. Hirashita for
their suggestions and discussions to carry out this work.
The author has made extensive use of NASA's Astrophysics Data System
Abstract Service (ADS).

\clearpage

\begin{deluxetable}{lccccccc}
\footnotesize
\tablecaption{Observational properties of Galactic H {\sc ii} regions}
\tablewidth{0pt}
\tablehead{\colhead{Object} & \colhead{$R$} & \colhead{$D$} &
 \colhead{$S_{\rm 5GHz}$} & \colhead{$F_{\rm IR}$} & \colhead{$T_{\rm e}$} &
 \colhead{$n_{\rm e}$} & \colhead{Class}\\  
\colhead{} & \colhead{(kpc)} &\colhead{(kpc)} & \colhead{(Jy)} &
 \colhead{($10^{-14}$ W cm$^{-2}$)} & \colhead{(K)} &
 \colhead{(cm$^{-3}$)} & \colhead{}\\
\colhead{(1)} & \colhead{(2)} & \colhead{(3)} & \colhead{(4)} & 
\colhead{(5)} & \colhead{(6)} & \colhead{(7)} & \colhead{(8)}\\}

 \startdata
 G1.13-0.11     & 0.2  & 8.5  & 1.6  & 2.1  & 5900 & 1000 & C \\
 W31B4          & 3.2  & 5.5  & 2.9  & 2.4  & 6800 & 1000 & C \\
 G23.95+0.15    & 4.3  & 5.2  & 1.78 & 2.3  & 6000 & 3500 & U \\
 G29.96-0.02    & 4.4  & 8.5  & 3.74 & 8.2  & 6100 & 1500 & C \\
 G25.38-0.18    & 5.3  & 11.5 & 2.5  & 3.9  & 6000 & 1600 & C \\
 G333.60-0.21   & 6.0  & 3.0  & 40   & 51   & 6200 & 4400 & U \\
 G45.45+0.06    & 6.5  & 8.3  & 2.5  & 3.8  & 7200 & 1000 & C \\
 W51e           & 6.7  & 7.3  & 11.5 & 13   & 7100 & 1600 & C \\
 NGC3576\tablenotemark{a}
                & 7.9  & 3.6  & 36.8 & 22.5 & 7500 & 6500 & U \\
 NGC3603P1      & 8.9  & 7.2  & 6.76 & 6.1  & 6900 & 1100 & C \\
 NGC3603P2      & 8.9  & 7.2  & 4.14 & 4.5  & 6900 & 1100 & C \\
 G298.22-0.34   & 9.9  & 10.5 & 22.2 & 7.4  & 8600 & 2200 & --- \\
 S156           & 10.2 & 3.4  & 1.4  & 1.1  & 6400 & 1000 & --- \\
 \enddata

\tablenotetext{a}{Cols.(4), (5), and (7) are averaged by two
 different observations.}

\tablecomments{Col. (1): Object name. Col.(2): Galactocentric
 radius. Col.(3): Distance from the Sun. Col.(4): Flux density at 5 GHz
 in each beam size which is selected to match the IR beam. Col.(5):
 Integrated IR flux in each beam size. The wavelength range is 18--160
 \micron. Col.(6): Electron temperature. Col.(7): Electron density from
 [OIII] doublets (52, 88 \micron). Col.(8): Classification in section
 3. 'C' denotes 'compact' H {\sc ii} regions, and 'U' denotes
 'ultra-compact' H {\sc ii} regions. All quantities are taken from
 \cite{sim95}.}

\end{deluxetable}

\clearpage

\begin{deluxetable}{lccccccc}
\footnotesize
\tablecaption{Determined properties of Galactic H {\sc ii} regions}
\tablewidth{0pt}
\tablehead{\colhead{Object} & \colhead{$N'_{\rm LC}$} & 
\colhead{$L_{\rm IR}$} & \colhead{$f$} & \colhead{$\log U$} & 
\colhead{$R_{\rm S}$} & \colhead{$r_{\rm i}$} & \colhead{$y_{\rm d}$}\\  
\colhead{} & \colhead{($10^{49} {\rm s}^{-1}$)} &\colhead{($10^6 L_\sun$)} & 
\colhead{} & \colhead{} & \colhead{(pc)} & \colhead{(pc)} & \colhead{}\\
\colhead{(1)} & \colhead{(2)} & \colhead{(3)} & \colhead{(4)} & 
\colhead{(5)} & \colhead{(6)} & \colhead{(7)} & \colhead{(8)}\\}

 \startdata
G1.13-0.11     & 1.3  & 0.59  & 0.44 & -1.94 & 0.84 & 0.64 & 0.18 \\
W31B4          & 0.93 & 0.28  & 0.63 & -2.07 & 0.69 & 0.59 & 0.38 \\
G23.95+0.15    & 0.54 & 0.24  & 0.45 & -1.89 & 0.27 & 0.21 & 0.24 \\
G29.96-0.02    & 3.0  & 2.3   & 0.28 & -1.70 & 1.00 & 0.65 & 0.15 \\
G25.38-0.18    & 3.7  & 2.0   & 0.38 & -1.70 & 0.92 & 0.67 & 0.29 \\
G333.60-0.21   & 4.0  & 1.8   & 0.45 & -1.58 & 0.46 & 0.35 & 0.45 \\
G45.45+0.06    & 1.8  & 1.0   & 0.36 & -1.91 & 1.05 & 0.75 & 0.14 \\
W51e           & 6.4  & 2.7   & 0.47 & -1.70 & 1.07 & 0.83 & 0.44 \\
NGC3576        & 4.8  & 1.1   & 0.77 & -1.62 & 0.33 & 0.30 & 0.75 \\
NGC3603P1      & 3.7  & 1.2   & 0.58 & -1.85 & 1.06 & 0.89 & 0.47 \\
NGC3603P2      & 2.3  & 0.91  & 0.49 & -1.90 & 0.95 & 0.75 & 0.34 \\
G298.22-0.34   & 23.  & 3.2   & 1.17 & -1.64 & 1.04 & 1.10 & 1.00 \\
S156           & 0.18 & 0.050 & 0.67 & -2.31 & 0.38 & 0.34 & 0.22 \\

 \enddata

\tablecomments{Col.(1): Object name. Col.(2): Production rate of
 ionizing photons estimated from 5GHz flux density. Col.(3):
 Total IR luminosity estimated from observed IR flux. The wavelength
 range is 8--1000 \micron. Col.(4): Determined fraction of ionizing
 photons. Mean uncertainty for individual value is estimated to be
 $\pm0.2$. Col.(5): Ionization parameter corrected from Lyman continuum
 extinction. Col.(6): Str{\"o}mgren radius corrected from Lyman
 continuum extinction in parsec unit. Col.(7): Ionized radius in parsec
 unit. Col.(8): Dust cavity's radius normalized by $R_{\rm S}$. The
 adopted $R_V$ is 5.5. Mean uncertainty for individual value is
 estimated to be $\pm0.2$.}

\end{deluxetable}

\clearpage

\begin{deluxetable}{lcccccccccc}
\footnotesize
\tablecaption{Mean properties of 'compact' and 'ultra-compact' H {\sc ii} 
 regions}
\tablewidth{0pt}
\tablehead{\colhead{Sample} & \colhead{Number} & \colhead{$T_{\rm e}$} 
& \colhead{$n_{\rm e}$} & \colhead{$N_{\rm LC}$}
& \colhead{$R_{\rm S}$} & \colhead{$r_{\rm i}$} & \colhead{$f$} 
& \colhead{$y_{\rm d}$} & \colhead{$r_{\rm d}/r_{\rm i}$} 
& \colhead{$r_{\rm d}$}\\  
\colhead{} & \colhead{} & \colhead{(K)} &\colhead{(cm$^{-3}$)} 
& \colhead{(10$^{49}$ cm$^{-1}$)} & \colhead{(pc)} 
& \colhead{(pc)} & \colhead{} & \colhead{} & \colhead{} 
& \colhead{(pc)}\\
\colhead{(1)} & \colhead{(2)} & \colhead{(3)} & \colhead{(4)} 
& \colhead{(5)} & \colhead{(6)} & \colhead{(7)} & \colhead{(8)}
& \colhead{(9)} & \colhead{(10)} & \colhead{(11)}\\}

 \startdata
compact       & 8 & 6600 & 1200 & 6.8 & 0.95 & 0.72 & 0.45 & 0.30 
& 0.38 & 0.28\\
 & & $\pm630$ & $\pm400$ & $\pm3.9$ & $\pm0.13$ & $\pm0.098$ & $\pm0.10$ 
& $\pm0.12$ & $\pm0.14$ & $\pm0.13$\\
ultra-compact & 3 & 6600 & 4800 & 5.4 & 0.35 & 0.29 & 0.56 & 0.48 
& 0.57 & 0.17 \\
 & & $\pm1100$ & $\pm1200$ & $\pm3.2$ & $\pm0.073$ & $\pm0.062$ &
 $\pm0.15$ & $\pm0.21$ & $\pm0.20$ & $\pm0.080$\\

 \enddata

\tablecomments{Col.(2): Number of sample regions. Col.(5): Intrinsic
 production rate of Lyman continuum photons Col.(7): Ionized
 radius in parsec unit. Col.(10): Ratio of the cavity's radius to
 ionized radius. Col.(11): Cavity's radius in parsec unit. We also show
 the standard deviations of all quantities.}

\end{deluxetable}

\clearpage

\begin{deluxetable}{lccc}
\footnotesize
\tablecaption{Mean dust cavity's radius for various $R_V$}
\tablewidth{0pt}
\tablehead{\colhead{Sample} & \colhead{} & \colhead{$R_V$} & \colhead{}\\
\colhead{} & \colhead{5.5} & \colhead{4.0} & \colhead{3.1}\\}

 \startdata
all           & 0.39 & 0.50 & 0.58\\
compact       & 0.30 & 0.43 & 0.52\\
ultra-compact & 0.48 & 0.57 & 0.63\\

 \enddata

\end{deluxetable}

\clearpage

\begin{figure}
\epsscale{.4}
\plotone{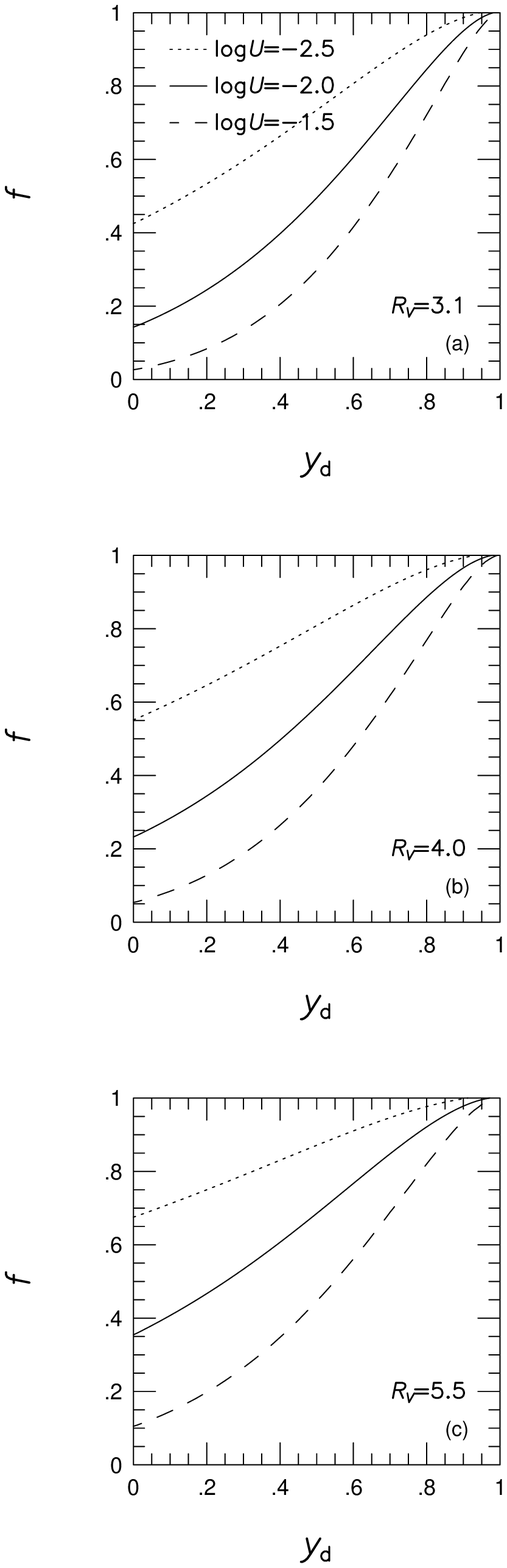}
\caption{Model calculation of ionizing photon fraction, $f$, as a
 function of the dust cavity's radius in unit of the Str{\"o}mgren
 radius, $y_{\rm d}$. For all panels, the dotted, solid, and dashed
 lines represent the cases of the logarithmic ionization
 parameter, $U$, of -2.5, -2.0, and -1.5, respectively. The electron
 temperature, $T_{\rm e}$, is set to be 6600 K. In panels (a),
 (b), and (c), we show the cases of the ratio of visual extinction to
 color excess, $R_V=3.1$, 4.0, and 5.5, respectively.}
\end{figure}

\begin{figure}
\epsscale{1.0}
\plotone{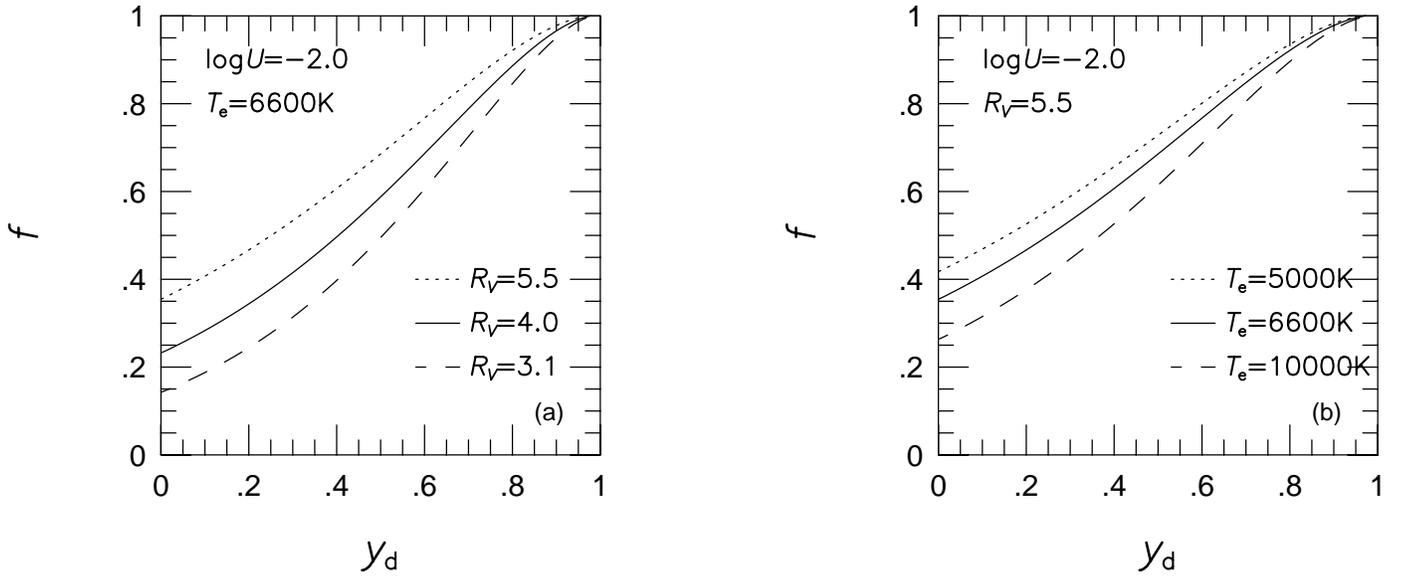}
\caption{Same as Figure 1. In panel (a), we show the variation of the 
 $f$-$y_{\rm d}$ relation of the case of $\log U=-2.0$ and $T_{\rm e}=6600$ 
 K for various $R_V$ values. In panel (b), we show the variation of the
 case of $\log U=-2.0$ and $R_V=5.5$ for various $T_{\rm e}$.}
\end{figure}

\begin{figure}
\epsscale{.7}
\plotone{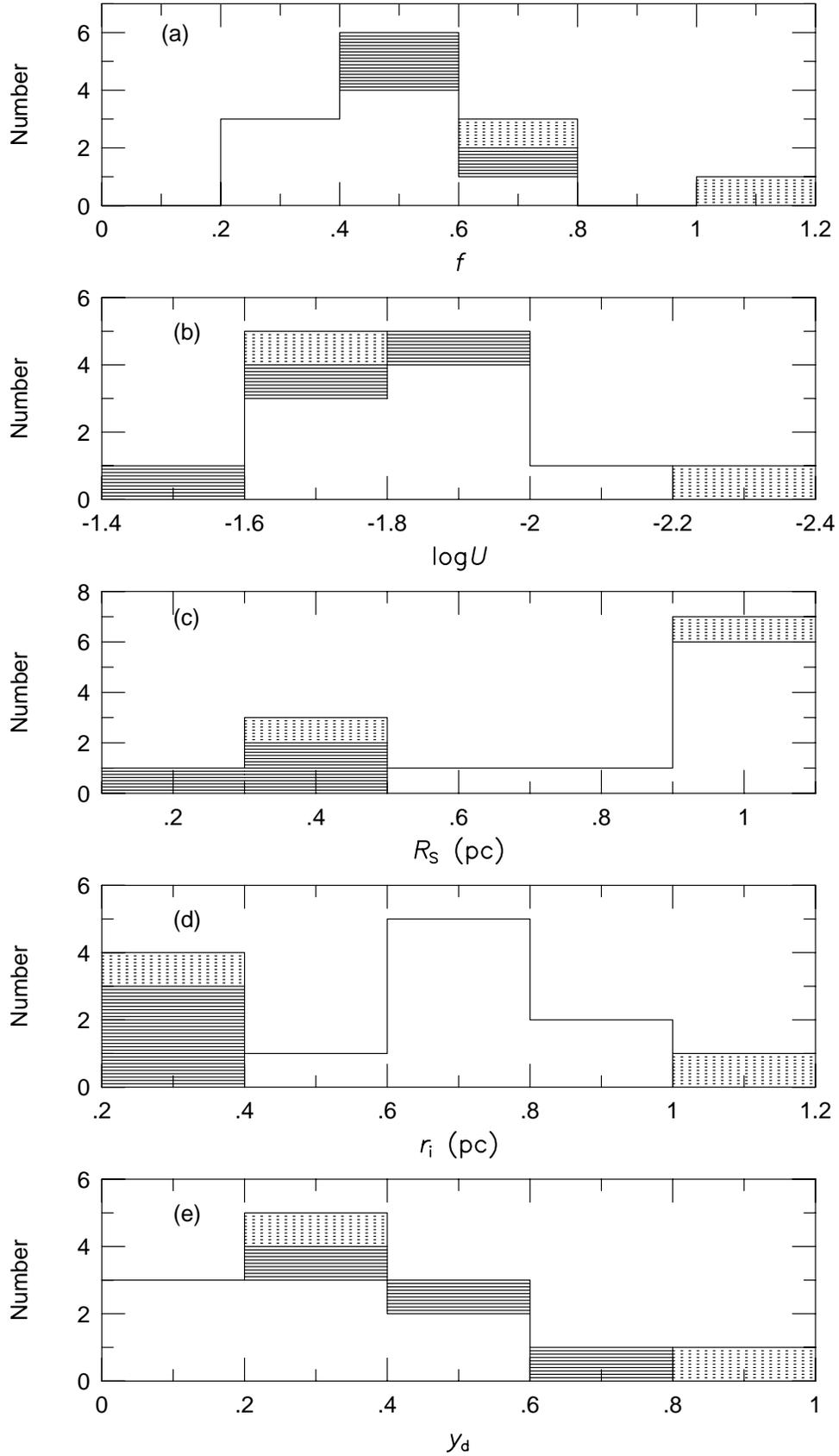}
\caption{Histograms of the determined quantities. (a) Ionizing photon 
 fraction. (b) logarithmic ionization parameter. (c) Str{\"o}mgran
 radius in parsec unit. (d) Ionized radius in parsec unit. (e) Dust
 cavity's radius in unit of the Str{\"o}mgren radius. The white
 histograms represents the 'compact' H {\sc ii} regions sample, and the
 shaded histograms represents the 'ultra-compact' sample. The dotted
 hisograms are two objects removed from further discussions.}
\end{figure}

\end{document}